\documentclass[sigconf,nonacm]{acmart}

\AtBeginDocument{%
  \providecommand\BibTeX{{%
    \normalfont B\kern-0.5em{\scshape i\kern-0.25em b}\kern-0.8em\TeX}}}

\usepackage{balance}
\usepackage{siunitx}
\usepackage{tabularx}
\begin{document}

\title{“Why SuaCode?”: Understanding African Students' Motivations for Taking a Smartphone-Based Online Coding Course}

\author{Michael Addo}
\affiliation{\institution{University of Ghana}}

\author{Nana Maryam Munagah}
\affiliation{\institution{Kwame Nkrumah University of Science and Technology}}

\author{Victor Kumbol}
\affiliation{
\institution{Kwame AI Inc.}
\institution{Charite Berlin}}

\author{Judith Uchidiuno}
\affiliation{\institution{Georgia Institute of Technology}}

\author{George Boateng}
\affiliation{
\institution{Kwame AI Inc.}
\institution{ETH Zurich}
}

\begin{abstract}
Computer programming MOOCs are instrumental in providing students with high-quality instruction in areas where there is limited access. They are especially beneficial to post-secondary African students as less than 1\% of them leave secondary school with fundamental coding skills. One strategy for increasing their efficacy for African students is to understand students’ motivation for enrolling. These insights can inform the design of MOOC content and assessments to align with students' interests. We administered an open-ended response survey to (self-identified) Africans enrolled in a smartphone-based online coding course (SuaCode). We analyzed a random sample of 450 (of 3000) responses using a grounded theory approach. We found that most African students (68.7\%) participated in SuaCode for intrinsic reasons such as improving themselves, learning with like-minded individuals, and gaining skills to help address societal issues. We discuss the implications of these findings in the design of programming MOOCs targeted at African students.
\end{abstract}



\maketitle

\section{Introduction}
There has been a sustained increase in the demand for post-secondary students with computer programming skills globally. However, African students lag behind the rest of the world with less than 1\% of them leaving secondary school with fundamental coding skills \cite{sap2016}. To address this demand, there has been a surge in non-traditional course offerings such as Massive Open Online Courses (MOOCs), coding boot camps, and online graduate degrees\cite{duncan2020}. One prominent offering in Africa is a coding workshop called Africa Code Week. The program has benefited over 4 million young Africans from 2015 to 2019 \cite{sap2016}, underscoring very high interest among Africans to learn to code. 

SuaCode Africa is a 2-month online programming course targeted at high school, university, and recent graduates living in Africa and the diaspora \cite{boateng2019}. It is a smartphone-based coding curriculum that covers basic programming concepts such as variables, conditionals, loops, functions, classes, objects, and arrays\cite{boateng2019}. The course uses the Processing programming language, an open-source, Java-based programming language, and students use the APDE Android app for writing and running their programs \cite{boateng2019}. These programs seek to bridge the gap in Africa by helping in the development of programming skills. However, developing good programming skills typically requires consistent practice, which students cannot sustain unless adequately motivated \cite{law2010}. Several studies have identified student motivation as a factor that affects engagement in online courses \cite{liyanagunawardena2013,sun2012}. However, there are limited studies evaluating motivations in programming MOOCs \cite{luik2019}, particularly in the African context.

In this study, we investigated students’ motivations for enrolling in SuaCode. Given the demand for coding programs and students' diverse cultural backgrounds, understanding students’ motivation can inform the design of MOOCs to better support and engage African students \cite{law2010}. The structure of the paper is as follows: in Section 2, we give a background on students' motivation for joining online coding courses like SuaCode. In Section 3, we describe the methods employed in collecting and analyzing our survey data. In Section 4, we present our results and discuss their implications. Finally, we share our study's limitations and our future work.

\section{Background and Related Work}
Massive Open Online Courses (MOOCs) are online courses that are designed for broad participation and open access to everyone, regardless of their educational background or geographic location \cite{psathas2018}. Several researchers have pointed out that motivation is a critical factor for student engagement in MOOCs as it affects their persistence and learning gains throughout the course \cite{Colquitt2000, liyanagunawardena2013, sun2012}. Motivation could be derived either intrinsically or extrinsically \cite{ryan2000}, and prior research studies have categorized students’ motivation using those classifications \cite{alraimi2015, white2014}. Intrinsic motivation is the learners’ interest and willingness to participate and elaborate on the course content to satisfy an internal need for self-improvement and actualization. Extrinsic motivation counts the learners’ willingness to engage in activities specifically to develop their professional career and partake in the prestige and reputation of the institutions delivering the MOOC \cite{psathas2018}. For example, Debara et al. conducted a study on student motivation and participation; they uncovered several intrinsic motivators such as personal interest, meeting future goals, and serving their values and beliefs \cite{debarba2016}.

Research studies show that extrinsic factors such as completing a certification in a coding MOOC had a lower value than a traditional course \cite{kizilcec2013}. Therefore, learners with intrinsic motivation may be more likely to persist through a course \cite{milligan2016, terras2015}. Interest in a topic (a feature of intrinsic motivation) has been identified as a motivator for participation in coding MOOCs by several studies \cite{bali2015, bonk2015, daza2013,debarba2016, kizilcec2013} \cite{white2014}. Learners participate in MOOCs because they find it fun and challenging \cite{kizilcec2013}. Our study considers other motivational components such as how social influences and demographic factors affect intrinsic and extrinsic motivators. We address the association between students' expressed intrinsic and extrinsic motivations and socio-demographic factors like age, gender, and educational level. Like other studies in this research area, we employ a mixed methods approach to analyze students’ motivation within SuaCode.

\section{Methodology}
\subsection{Developing the Coding Scheme}
We administered an open-ended response survey to understand students’ motivation for enrolling in SuaCode courses. We received over 3,000 responses from 2 cohorts of SuaCode courses. Two researchers coded the first 100 responses separately using a grounded theory approach. Most responses had multiple codes associated with them because nearly all respondents gave more than one reason for enrolling in SuaCode. From this analysis phase, we identified 19 distinct themes that motivated students to enroll after discussing and comparing the codes from the two raters with the entire team (Table 1, 2). 

Following this phase, the two raters coded another batch of 50 responses independently and found that they were in agreement with over 70\% of the students' responses. The codes were determined to be appropriate, and no new codes were discovered. A small number of responses indicated uncertainty or used unclear wording; these responses were coded as “Unclear”. A systematic random sampling technique was used to select 450 responses from over 3000 responses for the last round of coding completed together by two raters.

\begin{figure}
  \centering
  \includegraphics[width=\linewidth]{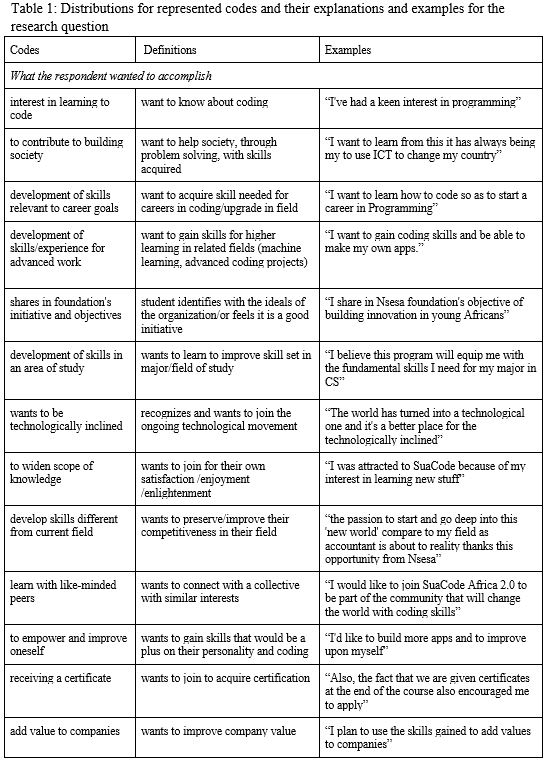}
  \Description{A table}
\label{fig:motivation1}
\end{figure}

\begin{figure}
  \centering
  \includegraphics[width=\linewidth]{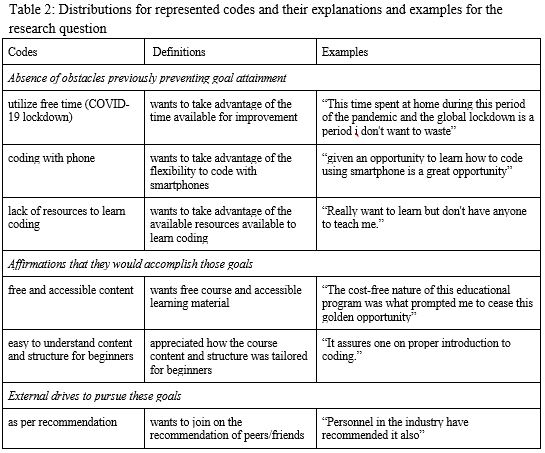}
  \Description{A table}
\label{fig:motivation2}
\end{figure}

\section{Results and Discussion}
\begin{figure}
    \centering
    \includegraphics[width=\linewidth]{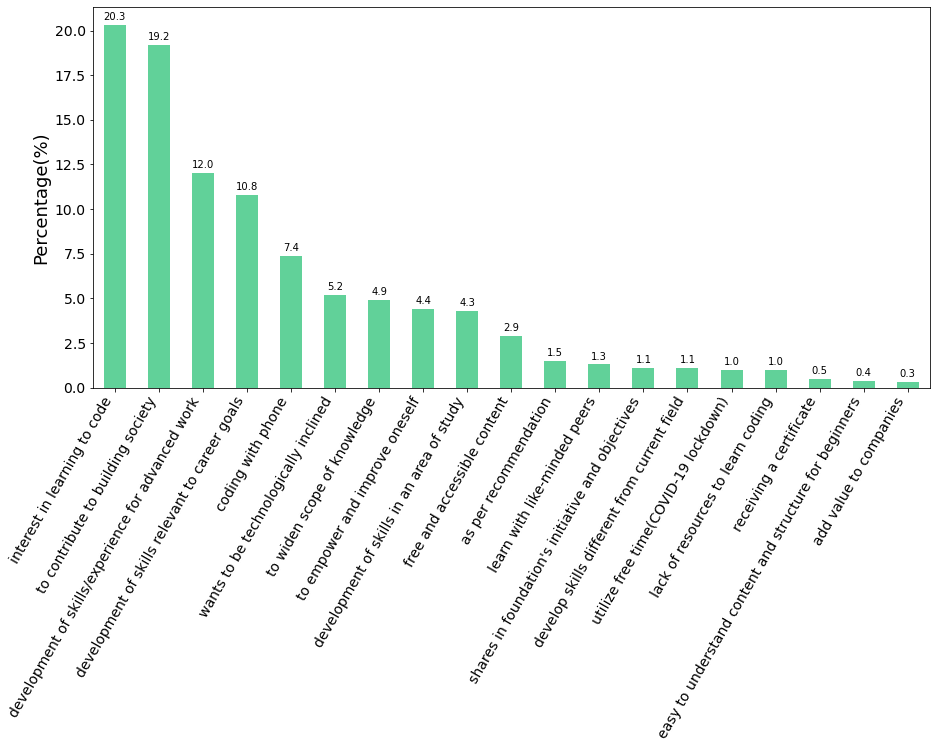}
    \Description{A bar chart}

    \caption{A frequency distribution of codes/concepts developed after a consensus for the first 100 responses of the “Why do SuaCoders want to take part in SuaCode courses?” dataset was coded by two raters.}
    \label{fig:motivation_fig}
\end{figure}

Figure \ref{fig:motivation_fig} shows that 20.3\% of the students from the 450 analyzed enrolled because they are interested in learning to code, 19.2\% applied so they could impact and contribute to building and developing society and 12.0\% want to develop new skills, and experience for advanced work.

Most participants identified as male (69.64\%), within the 20-29 age bracket (70.04\%). The majority of them were in college/university education levels (76.73\%). Most participants expressed a diversity of motivations for enrolling in SuaCode: some were extrinsic e.g., tied to an external, objective value such as the development of skills relevant to their career goals, developing skills/experience for advanced work, developing skills different from the current field, adding value to companies, receiving a certificate, etc. Others were more intrinsic to the individual such as interest in learning to code, contributing to building society, developing skills in an area of study, learning with like-minded peers, widening their scope of knowledge, empowering and improving oneself, etc. Extrinsic motivators comprised 31.3 \% of codes assigned, while intrinsic motivators comprised 68.7\%. This is in line with current literature on motivations in MOOCs \cite{bonk2015, duncan2020}, as participants of SuaCode are intrinsically motivated learners.

We found that the code "free, understandable, and accessible content" (4\%) ranked lower compared to other motivation-based studies. White et al. in an Estonian study \cite{white2014}, found that students cited the cost and accessibility of online programs as one of the three most marked reasons for choosing to join a MOOC. This trend suggests that providing no-cost and accessible learning content may not be a strong enough motivator for African students to enroll in MOOCs without accounting for other factors. Although SuaCode has high adoption due to its accessibility via smartphones, student responses suggest that it is their interest in the program’s theme and applicability to their goals that primarily drive their enrollment.

\section{Limitations, Future Work and Conclusion}
This poster presents a preliminary analysis of African students’ survey responses on their motivation to engage in a smartphone-based computer programming MOOC called SuaCode. We found that most students were motivated to take this course for intrinsic reasons and that free and accessible content was not a strong enough motivator for their enrollment. 

We are currently analyzing the full dataset for a longer-form paper submission investigating different questions such as how men and women may be motivated differently, how student motivations differ across different regions of Africa, and how younger students may be motivated differently from older and working professionals. More importantly, we are investigating how these different intrinsic and extrinsic motivational factors influence students’ ability to persist through the course, and what motivational factors matter the most for those that achieve the highest learning gains.  Analyzing all 3000+ responses will likely uncover other codes that influence the ways we understand African students’ motivation in MOOCs, and how to better design computer programming smartphone-based MOOCs to support them. 

\begin{acks}
We are grateful to the Processing Foundation for supporting this work via the Processing Foundation Fellowship program 
\end{acks}

\balance
\bibliographystyle{ACM-Reference-Format}
\bibliography{refs}

\end{document}